\documentclass[reprint,superscriptaddress,showkeys]{revtex4-1}
\pdfoutput=1
\usepackage[utf8]{inputenc}
\usepackage{amsmath}
\usepackage{amssymb}
\usepackage{amsfonts}
\usepackage{graphicx}
\usepackage{psfrag}
\usepackage{siunitx}
\usepackage{color}

\newcommand{\Tfluc}{T^\mathrm{fluc}}
\newcommand{\Tm}{T^\mathrm{m}}
\newcommand{\Tavg}{T^\mathrm{avg}}

\begin{document}

\title{Extended Nyquist formula for a resistance subject to a heat flow}

\author{Benjamin Monnet}
\author{Sergio Ciliberto}
\author{Ludovic Bellon}
\email{Corresponding author : ludovic.bellon@ens-lyon.fr}
\affiliation{Univ Lyon, Ens de Lyon, Univ Claude Bernard Lyon 1, CNRS, Laboratoire de Physique, F-69342 Lyon, France}

\date{\today}

\begin{abstract}
The Nyquist formula quantifies the thermal noise driven fluctuations of voltage across a resistance in equilibrium. We deal here with the case of a resistance driven out of equilibrium by putting it in contact with two thermostats at different temperatures. We reach a non-equilibrium steady state where a heat flux is flowing through the resistance. Our measurements demonstrate anyway that a simple extension of the Nyquist formula to the non uniform temperature field describes with an excellent precision the thermal noise. For a metallic ohmic material, the fluctuations are actually equivalent to those of a resistance in equilibrium with a single thermostat at the mean temperature between the hot and cold sources. 
\end{abstract}

\maketitle

\section {Introduction}

With the Fluctuation-Dissipation Theorem (FDT), statistical physics offers a powerful tool to describe the fluctuations of an observable of a system in equilibrium~\cite{Callen-1951}. For example, applications of the FDT to the Brownian motion of micrometer sized systems include micro-rheology measurements with optical tweezers~\cite{Mizuno-2008,Pesce-2009}, or the calibration of the stiffness of atomic force microscopy probes~\cite{Butt-1995}. For electrical systems, the FDT is expressed by the celebrated Nyquist formula~\cite{Nyquist-1928}, describing the Johnson voltage noise~\cite{Johnson-1928} across a resistance $R$ at temperature $T$:
\begin{equation} \label{eq:Nyquist}
S_V = \frac{\langle V^2 \rangle}{\Delta f} = 4 k_B T R
\end{equation}
with $S_V$ the Power Spectrum Density (PSD) of the voltage $V$ across the resistance, $\Delta f$ the frequency bandwidth, $k_B$ the Boltzmann constant and $\langle . \rangle$ stands for the statistical average.

In non equilibrium situations however, such a relation between fluctuations and dissipation is not granted, and excess noise is usually expected to be observed with respect to an equilibrium state~\cite{Conti-2013,Conti-2014,Li-1994,Cugliandolo-1997,Li-1998,Grigera-1999,Bellon-2001,Berthier-2002,Herisson-2002,Grenard-2008,Monchaux-2008,Marconi-2008,Loi-2011,Santamaria-Holek-2011,Cugliandolo-2011,Kasas-2015,Dieterich-2015}. We will focus here on the case where the system is in a Non-Equilibrium Steady State (NESS), because it is submitted to a constant heat flux. This is the case for example of a conductor submitted to a large temperature gradient, or of a resonator whose extremity are receiving an external radiation. In this latter case, contradictory observations have been made: in a experiment by L. Conti and collaborators~\cite{Conti-2013,Conti-2014}, an excess of thermal noise has been measured, while in a similar experiment in our group~\cite{Geitner-2017}, lower fluctuations than those expected from the system average temperature have been observed.

To give further insights into these puzzling physical phenomena, we explore in this article the thermal fluctuation in such a NESS, focusing on the simple example of voltage noise across an electrical resistance. In a first part, we present the simple extension of the Nyquist formula that could be applied to this system. In the following parts, we describe the experimental setup and the results of the thermal noise measurements. The last section concludes this work by a discussion on the results.

\section{Extended Nyquist formula}

To introduce our approach, let us first discuss the case of two resistances $R_1$ and $R_2$ connected in series, each being in equilibrium with a different thermostat at temperature $T_1$ and $T_2$ respectively. The PSD of the voltage noise across each resistance is described by the Nyquist formula (eq.~\ref{eq:Nyquist}). The voltage across the total resistance $R=R_1+R_2$ is $V=V_1+V_2$, where we neglect the resistance of the conductor linking the two dipoles. Voltage fluctuations $V_1$ and $V_2$ are statistically uncorrelated, thus the measured PSD should write:
\begin{equation} \label{eq:Nyquist1+2}
S_V = \frac{\langle (V_1+V_2)^2 \rangle}{\Delta f} = S_{V_1} + S_{V_2} = 4 k_B (T_1 R_1 + T_2 R_2)
\end{equation}
The voltage noise is thus proportional to the sum of the products between temperature and resistance. The fact that there are two thermostats is decoupled here from the thermal noise analysis, since the heat flux between them only takes place in a conductor with negligible resistance. If we consider a similar case where $N$ resistances $R_n$ in equilibrium at temperature $T_n$ are connected in series, we will simply derive the following formula:
\begin{equation} \label{eq:NyquistN}
S_V = 4 k_B \sum_{n=1}^{N} T_n R_n
\end{equation}

Let us now consider the continuum limit of this approach, where each segment $dx$ of the resistor is at a local temperature $T(x)$. Under this hypothesis of local equilibrium, a possible extension of eq.~\ref{eq:NyquistN} is:
\begin{equation} \label{eq:extNyquist}
S_V = 4 k_B \int_{x=0}^{L} T(x) \rho\big(T(x)\big) dx
\end{equation}
where we suppose that the resistance is distributed in one dimension (along $x$, from $x=0$ to $x=L$), with a linear resistivity $\rho(T)$: 
\begin{equation}
R = \int_{x=0}^{L} \rho\big(T(x)\big)  dx
\end{equation}
This description is however not granted, since now each segment is submitted to a heat flux, thus equilibrium laws may not apply to this NESS. The goal of our experiments is to test this extended Nyquist formula, using a resistor in contact with two different thermostats at each of its extremities.

In the following, instead of comparing directly the measured PSD $S_V$ to eq.~\ref{eq:extNyquist}, we will characterise the amplitude of thermal noise by $\Tfluc$, defined with:
\begin{equation} \label{eq:Tflucdef}
\Tfluc := \frac{S_V}{4 k_B R}
\end{equation}
$\Tfluc $ is the effective temperature one would infer from measuring both the amplitude of the thermal noise and the value of the resistance, regardless of its equilibrium state. At equilibrium, this should be the thermostat temperature. Out of equilibrium, if the extended Nyquist formula is valid, we expect that
\begin{equation} \label{eq:Tflucext}
\Tfluc = \frac{1}{R} \int_{x=0}^{L} T(x) \rho\big(T(x)\big)  dx
\end{equation}
In other words, we expect $\Tfluc$ to be the temperature field averaged with a weight proportional to the local resistivity. Although the systems are very different (purely dissipative versus resonant system), this formula extends to electrical observables our work on the mechanical thermal noise of a micro-cantilever in a NESS~\cite{Geitner-2017}. K. Komori and collaborators came independently to a similar expression for the thermal noise of a generic mechanical system in a NESS~\cite{Komori-2018}.

\section{Measuring the thermal noise of a resistor in a thermal gradient}

In the experiments, we measure the voltage fluctuations using a home made ultra low noise amplifier \cite{Cannata-2009,Ciliberto-2013}, featuring a voltage noise of only $\SI{E-18}{V^2/Hz}$, flat down to $\SI{10}{Hz}$ (see Fig.~\ref{fig:PSD}). This is equivalent to the thermal noise of a $\SI{60}{\Omega}$ resistance at room temperature. Its current noise is negligible in our experiments (order of magnitude: $\SI{E-30}{A^2/Hz}$). Though we will subtract the amplifier noise spectrum from our measured PSD, having a reasonable signal to noise ratio implies using resistances greater than a few $\SI{100}{\Omega}$.

\begin{figure}[htbp] 
\begin{center}
\includegraphics[width=85mm]{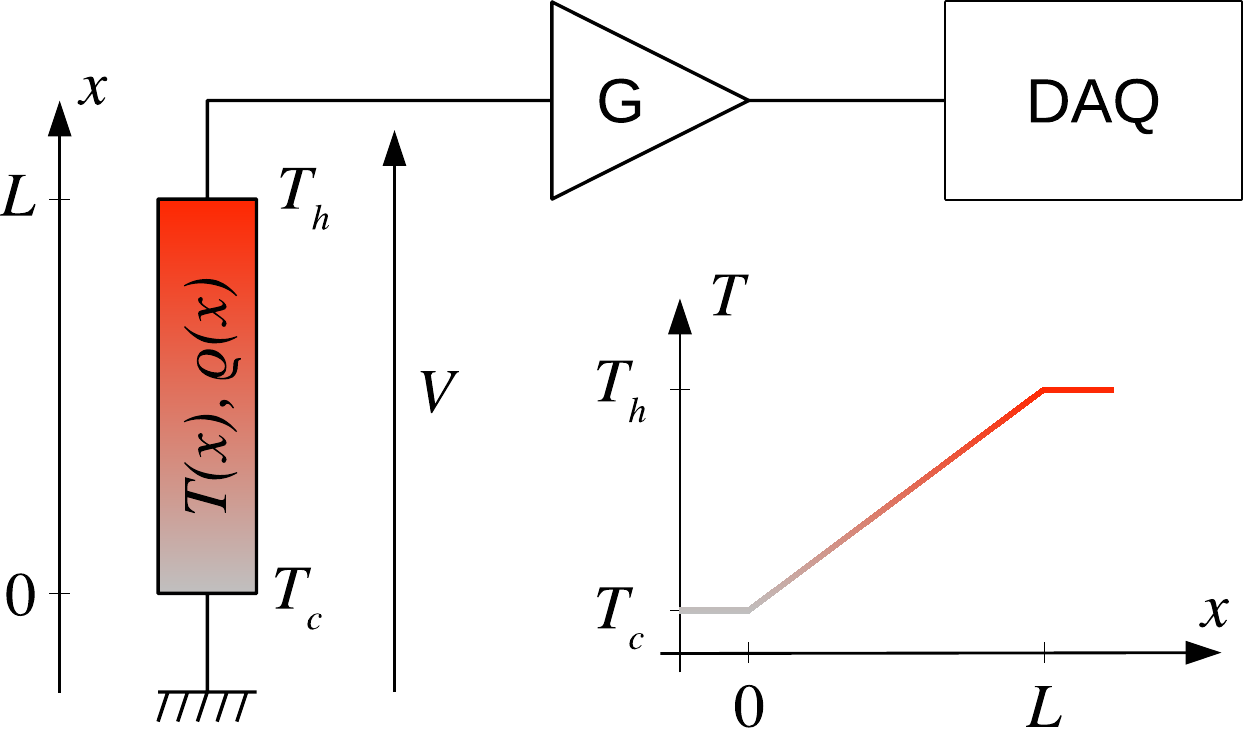}
\caption{Principle of the experiment: we measure the thermal noise driven voltage fluctuation across a resistance submitted to a temperature difference. The signal is amplified with a low noise voltage amplifier (gain $G=\num{e3}$) and acquired at $\SI{6}{kHz}$ with a 24 bits data acquisition card (NI PXI 4461).} \label{fig:IdealSketch}
\end{center}
\end{figure}

Ideally, we would like to mesure the voltage fluctuations across a resistance as sketched in Fig.\ref{fig:IdealSketch}. However, to avoid having local thermostats along the resistance (which would bring us close to the equilibrium case of eq.~\ref{eq:NyquistN}), we would like to avoid any material not part of the resistance to be in contact with it. We may not use a standard film resistor for instance, where most of the dipole is made of an insulating ceramic. We therefore choose to work with a resistive wire, where electrical and thermal conductivities are mainly due to the electronic transport. The material, length $L$ and diameter $D$ of this resistive wire need to accommodate the following criteria:
\begin{enumerate}
\item The total resistance $R = 4 \rho_v L / \pi D^2$ is greater than a few $\SI{100}{\Omega}$ (with $\rho_v$ the volume resistivity).
\item The heat flux by conduction within the resistance is much greater than that due to radiation.
\item The heat flux by conduction within the resistance is much greater than that with the surrounding atmosphere.
\end{enumerate}

\subsection{Resistance criterium}
To meet the first criterium, we need a long and thin wire of high electrical resistivity. If we choose for example Ni-Fe alloys (among the worst metallic conductors), their bulk resistivity is around $\rho_v=\SI{5e-7}{\Omega m}$. Thus a wire of diameter $D=\SI{25}{\mu m}$ will need to be at least $L=\SI{30}{cm}$ long to reach $R=\SI{300}{\Omega}$.

\subsection{Negligible radiation criterium}

Let us estimate the balance between conduction and radiation in a cylindrical wire bridging a cold thermostat at temperature $T_c$ with a hot one at temperature $T_h=T_c+\Delta T$. The equation governing the temperature field is derived by an analysis along the wire of the heat fluxes $J_\lambda$ by conduction and $J_\sigma$ emitted by radiation. The first follows the Fourier law:
\begin{equation}
J_\lambda = - \lambda \frac{\pi}{4} D^2 \partial_x T
\end{equation}
with $\lambda$ the thermal conductivity. We neglect any variation of $\lambda$ with $T$ for this estimation of the effect of radiation. As for $J_\sigma$, we suppose that the resistor is exchanging thermal photons with a half space at temperature $T_c$, and a second half space at temperature $T_h$. For a element of length $dx$, we thus write
\begin{equation}
J_\sigma = e \sigma (T^4-T_\sigma^4) \pi D dx
\end{equation}
with $\sigma=\SI{5.67E-8}{Wm^{-2}K^{-4}}$ the Stephan-Boltzmann constant, $e$ the emissivity of the material, and $T_\sigma=(T_c^4/2+T_h^4/2)^{1/4}$. The temperature profile is a balance between the fluxes for the element $dx$:
\begin{align}
\lambda \frac{\pi}{4} D^2 \partial_x^2 T & = e \sigma (T^4-T_\sigma^4) \pi D \\
\frac{1}{T_\sigma} \partial_x^2 T & = \frac{4 e \sigma T_\sigma^3}{\lambda D}(\frac{T^4}{T_\sigma^4}-1)
\end{align}
We define the characteristic length $L_\sigma$ by
\begin{equation}
L_\sigma=\sqrt{\frac{\lambda D}{4 e \sigma T_\sigma^3}}
\end{equation}
Normalising temperatures by $T_\sigma$ ($\theta=T/T_\sigma$) and positions by $L$ ($X=x/L$), we end up with the following adimensional equation:
\begin{equation} \label{eq:theta"}
\partial_X^2 \theta = \epsilon^2 (\theta^4 -1)
\end{equation}
with $\epsilon = L/L_\sigma$. Heat conduction will thus dominate if $\epsilon\ll 1$, and radiation will dictate the temperature field otherwise.

The second criterium thus favours long and thin wires, of low thermal conductivity. Given the Wiedemann–Franz law for metallic material stating that thermal and electrical conductivities are proportional~\cite{Franz-1853,Jones-1985}, the two first criteria are exactly antinomic ! For Ni-Fe alloys for example, the thermal conductivity is around $\lambda=\SI{20}{W m^{-1} K^{-1}}$, using $e=1$ (worst case corresponding to black body emission), $T_\sigma=\SI{350}{K}$ and $D=\SI{25}{\mu m}$, we compute $L_\sigma=\SI{7}{mm}$, which is much smaller than $L=\SI{30}{cm}$ imposed by the first criterium. A naive geometry is thus not compatible with the phenomenon we want to probe, and we must look at some alternative design.

\begin{figure}[htbp] 
\begin{center}
\includegraphics[width=85mm]{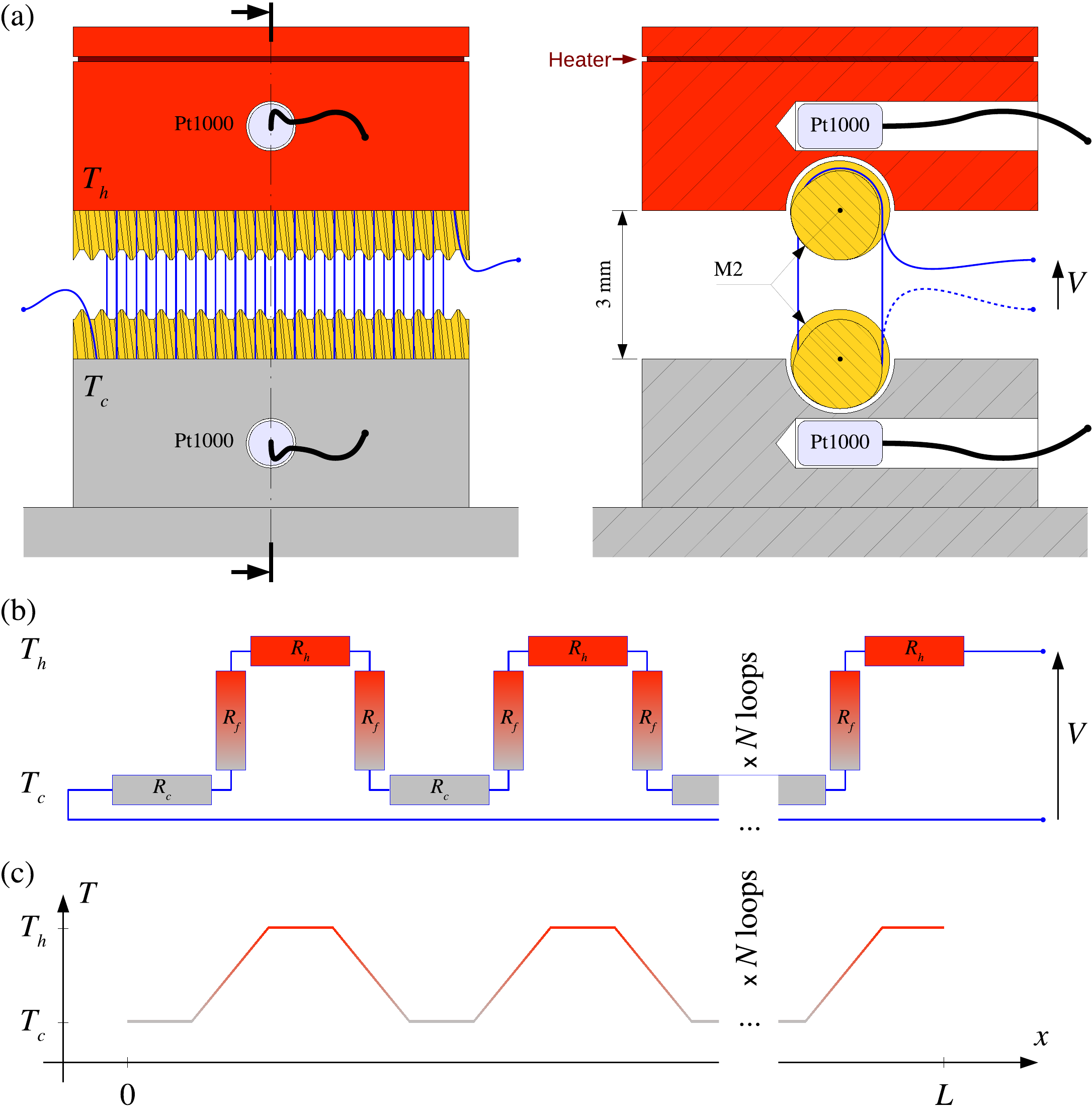}
\caption{Sketch of the experimental setup (a), equivalent electrical circuit (b) and temperature profile in the resistance (c). The resistance is a resistive wire (in blue) wrapped about 30 times around 2 brass screws (in yellow), which are part of the 2 thermostats (in gray at $T_0$ and red at $T_0+\Delta T$). Pt1000 thermistances are placed inside the thermostats to measure their temperature. A foil heater (in brown) is used to tune $\Delta T$. The voltage $V$ is measured across the whole resistance, but their thermal fluxes are in parallel in each free standing part of the wire.} \label{fig:Sketch}
\end{center}
\end{figure}

High thermal conductivity and strong electrical resistance are not achievable with a linear geometry. Our design compensates this incompatibility by using small resistances in parallel for heat fluxes but in series for the electrical resistance. This is achieved by wrapping the resistive wire around two cylindrical thermostats (M2 brass screws, $2r \approx \SI{1.6}{mm}$ in diameter), only $l=\SI{3}{mm}$ appart (axis distance). With $N\approx30$ loops, we end up with the required $\SI{30}{cm}$ length. To avoid electrical shortcuts, the resistive wire is coated with a thin insulating layer. Thermal conducting grease on the brass screws ensure good thermal contact between the resistance and the thermostat. The resulting electrical system, sketched in Fig.~\ref{fig:Sketch}, is the series of resistances $R = N (R_c+R_f+R_h+R_f)$, where the three subscripts stand for the part of the resistive wire in contact with the cold thermostat ($c$), the hot one ($h$), and the free standing part bridging the two ($f$). $R_c$ and $R_h$ share the same length $l_c=l_h=\pi r \approx \SI{2.5}{mm}$, comparable to that of $R_f$ : $l_f = l =\SI{3}{mm}$.

\begin{figure}[htbp] 
\begin{center}
\includegraphics{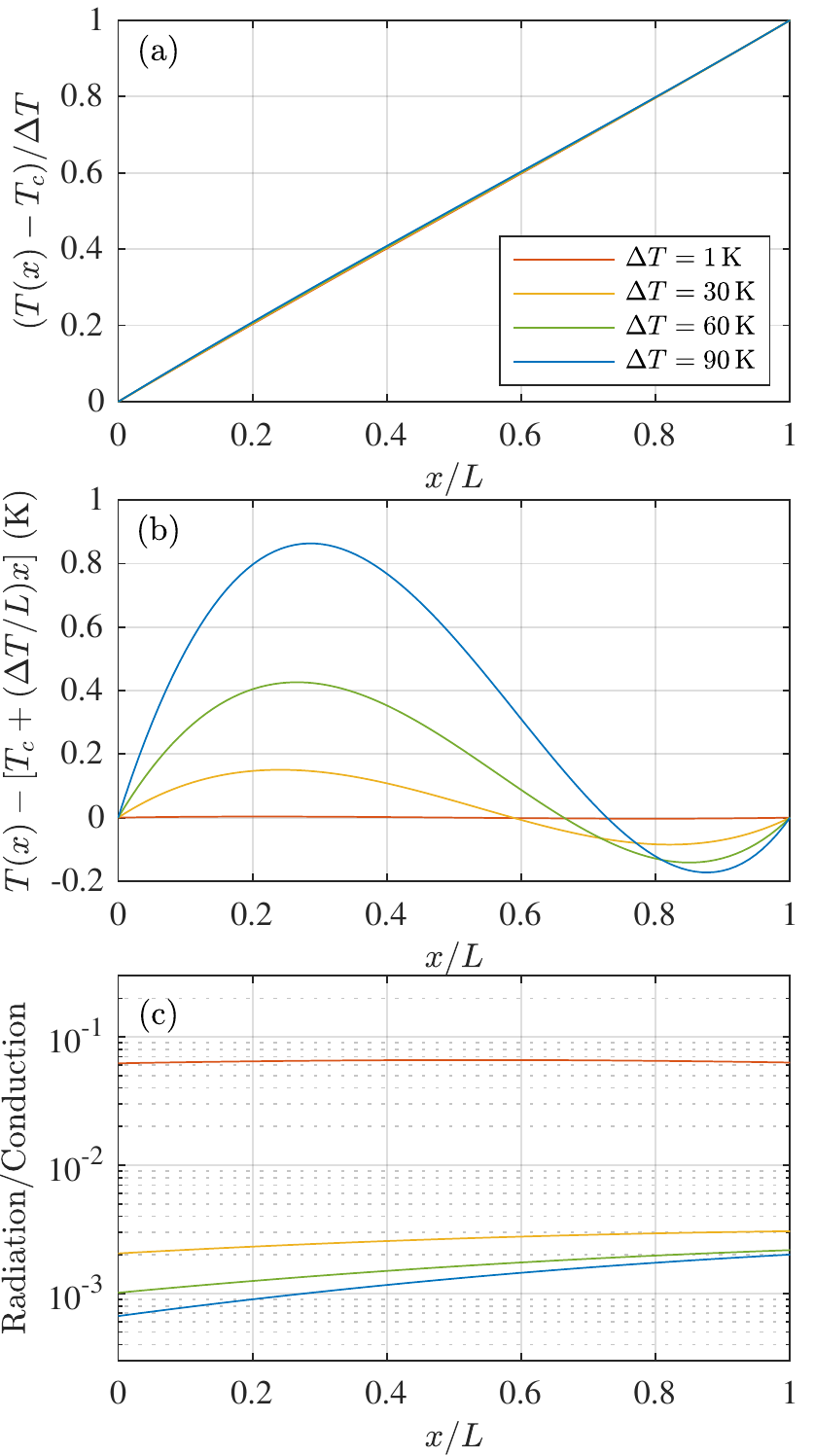}
\caption{Simulated temperature profile (a), deviation to the purely conductive case (b), and ratio between radiation and conduction heat fluxes (c), for various $\Delta T$, with eq. \ref{eq:theta"} using the following parameters: $l=\SI{3}{mm}$, $D=\SI{25}{\mu m}$, $\lambda=\SI{20}{W m^{-1} K^{-1}}$, $e=1$, and $T_c=\SI{295}{K}$. In this specific geometry, radiation is negligible versus conduction.} \label{fig:sim}
\end{center}
\end{figure}

As illustrated in Fig.~\ref{fig:sim}, numerically solving eq.~\ref{eq:theta"} for $R_f$ in this specific case ($l=\SI{3}{mm}$, $D=\SI{25}{\mu m}$, $\lambda=\SI{20}{W m^{-1} K^{-1}}$, $e=1$) shows that deviations to the purely conductive temperature profile are below $\SI{1}{\%}$. To estimate if the thermal photon bath can be considered as a local thermostat, we can also compare the conduction heat flux per unit surface $j_\lambda= - \lambda \partial_x T$ to that of emitted radiations $j_\sigma = e \sigma T^4$ . In figure Fig.~\ref{fig:sim}(c), we observe that this ratio $j_\sigma/j_\lambda$ is at most a few percent, and even much smaller for the largest $\Delta T$.

As for thermal noise measurements, each of the $N$  loops around the thermostats has an equivalent contribution both for the total resistance ($R=N R_\mathrm{loop}$) and the total voltage PSD ($S_V = N S_{V_\mathrm{loop}}$), hence $\Tfluc$ defined by eq.~\ref{eq:Tflucdef} is the same for one loop or the full resistance. Finally, the extended Nyquist Formula in this geometry is equivalent to:
\begin{align}
\Tfluc	& = \frac{1}{R_\mathrm{loop}} \int_{x=0}^{L_\mathrm{loop}} T(x) \rho\big(T(x)\big)  dx \\
		& = \frac{R_c T_c+ R_h T_h +  2 R_f \Tfluc_f}{R_c + R_h + 2 R_f} \label{eq:Tflucloop}
\end{align}
with 
\begin{equation} \label{eq:Tflucf}
\Tfluc_f = \frac{1}{R_f}\int_{x=0}^{l} T(x) \rho\big(T(x)\big)  dx
\end{equation}
To the first order in $\Delta T$, we can neglect the variation of $\rho$ and $\lambda$ with temperature, hence the temperature profile is linear ($T(x) = T_c + \Delta T x / L$) and eqs.~\ref{eq:Tflucloop} and \ref{eq:Tflucf} simply becomes:
\begin{equation} \label{eq:TflucLin}
\Tfluc = \Tfluc_f = T_c + \frac{\Delta T}{2}
\end{equation}

\subsection{Negligible convection criterium}

We now estimate the balance between conduction inside the resistive wire and heat flowing to or from the surrounding atmosphere. This air layer is confined in our setup between two horizontal plates (the thermostats), the hotter one being on top. This configuration is stable with respect to convection, the temperature will thus be linear in space in the medium around the resistance. As detailed in the previous paragraph, to the first order in $\Delta T$ the temperature profile due to conduction is also linear in space inside the resistance. The temperatures being equal at the same position, no net heat exchange with the environment is thus happening.

Let us anyway model this heat exchange $J_a$ if a difference were to appear between the resistance temperature $T(x)$  and the air temperature $T_a$, by introducing the heat exchange coefficient $h$~\cite{Bergman-2011}: 
\begin{equation}
J_a = h (T-T_a) \pi D dx
\end{equation}
where we considered a small element of length $dx$, as for radiation. For free convection at the vertical surface of a solid in contact with air around room temperature, the order of magnitude is $h\sim\SI{10}{W m^{-2} K^{-1}}$~\cite{Bergman-2011}. For a stationary solution, the balance between the fluxes for the element $dx$ writes:
\begin{align}
\lambda \frac{\pi}{4} D^2 \partial_x^2 T & = h (T-T_a) \pi D \\
\partial_x^2 T & = \frac{4 h}{\lambda D}(T-T_a)
\end{align}
The characteristic length $L_a$ for this equation is
\begin{equation}
L_a=\sqrt{\frac{\lambda D}{4 h}}\sim\SI{4}{mm}
\end{equation}
If significative temperature differences between the air and the resistive wire were to be present, the characteristic length $L_a$ over which the profile $T(x)$ would be impacted is thus larger than the wire actual free standing length $l=\SI{3}{mm}$. Again, thanks to the sample design, heat exchanges with air can thus be neglected, and the resistive wire environment cannot be considered as a local thermostat.

\section{Measurement results}

\subsection{Experimental setup and measurement of temperature coefficient}

Guided by the design considerations above, our sample is made of a wire of Ni70Fe30~\cite{GoodFellow}, of diameter $D=\SI{25}{\mu m}$ and length around $L=\SI{30}{cm}$. The wire is electrically insulated by a thin Polyimide layer (thickness $\SI{3.5}{\mu m}$). Its total resistance at room temperature is $R=\SI{295}{\Omega}$. It is wrapped approximately 30 times around the two brass screws that act as thermostats. Each screw is half buried (along its length) in a thick aluminium plate whose temperature is controlled: the low temperature thermostat is in contact with a one square meter large metallic plate at room temperature, and the high temperature thermostat is heated by a foil electrical resistance (sketch on Fig.~\ref{fig:Sketch}). Pt1000 temperature sensors are placed inside the aluminium plates, as close as possible to the brass screws, to measure $T_c$ and $T_h=T_c+\Delta T$. We checked that the temperature was uniform within $\SI{0.2}{K}$ in each thermostat. The mid point between these two readings gives an estimation of the average temperature of the resistance:
\begin{equation} \label{eq:defTm}
\Tm:=\frac{T_c+T_h}{2}=T_c+\frac{1}{2}\Delta T
\end{equation}

\begin{figure}[htbp] 
\begin{center}
\includegraphics{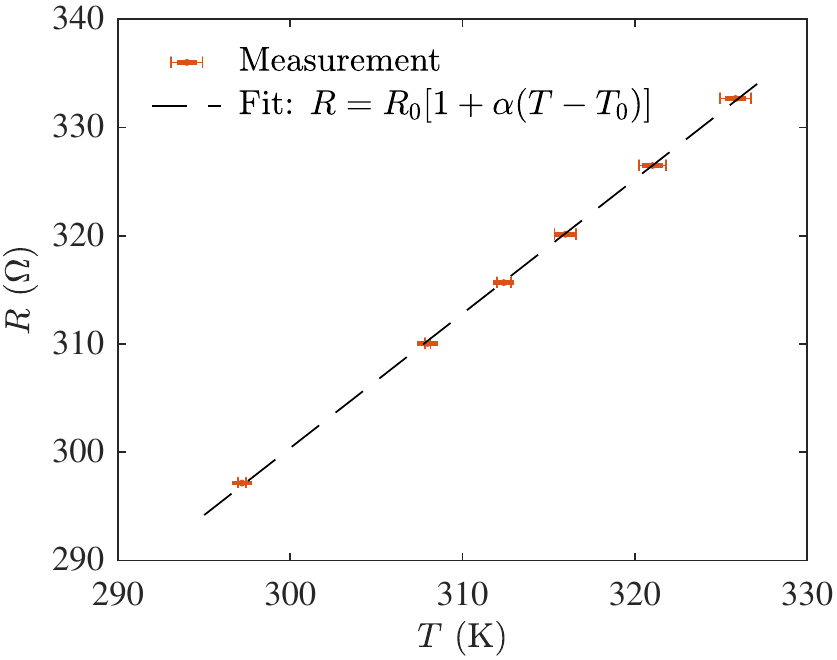}
\caption{Resistance of the sample as a function of its uniform temperature. The dependency is linear: $R=R_0 [1+\alpha (T-T_0)]$ with $\alpha=\SI{4.1e-3}{K^{-1}}$ and $R_0=\SI{300.4}{\Omega}$ at $T_0=\SI{300}{K}$.} \label{fig:R0alpha}
\end{center}
\end{figure}

In a first set of experiments, we characterise the variation of resistance (hence resistivity) with temperature. To this aim, we thermally shortcut the two thermostats and insulate them from the environment to reach a uniform temperature for all the system. All the readings are uniform within $\Delta T < \SI{1.7}{K}$, as displayed with the horizontal error bars in Fig.~\ref{fig:R0alpha}. We fit the measurement with a linear law:
\begin{equation}
R=R_0 [1+\alpha (T-T_0)]
\end{equation}
with $\alpha=\SI{4.1e-3}{K^{-1}}$ and $R_0=\SI{300.4}{\Omega}$ at $T_0=\SI{300}{K}$. This law applies as well to resistivity: $\rho=\rho_0 [1+\alpha (T-T_0)]$.

Using this linear dependency, we can infer an average value of the temperature of the resistance from its value when $\Delta T\neq 0$:
let us define $\Tavg$ by
\begin{equation} \label{eq:Tavgdef}
\Tavg := T_0+\frac{1}{\alpha}(\frac{R}{R_0}-1)
\end{equation}
Straightforward calculations show indeed that:
\begin{equation}
\Tavg = \frac{1}{L}\int_{x=0}^{L} T(x) dx
\end{equation}

To the first order in $\Delta T$, $T(x)$ being linear in space, we have $\Tavg=\Tm$ (with $\Tm$ defined by eq.~\ref{eq:defTm}). In the measurements, we will thus be able to compare the amplitude of thermal noise, measured with $\Tfluc$, with two independent measurements of the mean temperature of the system ($\Tm$ and $\Tavg$).

\subsection{Thermal noise}
\begin{figure}[htbp] 
\begin{center}
\includegraphics{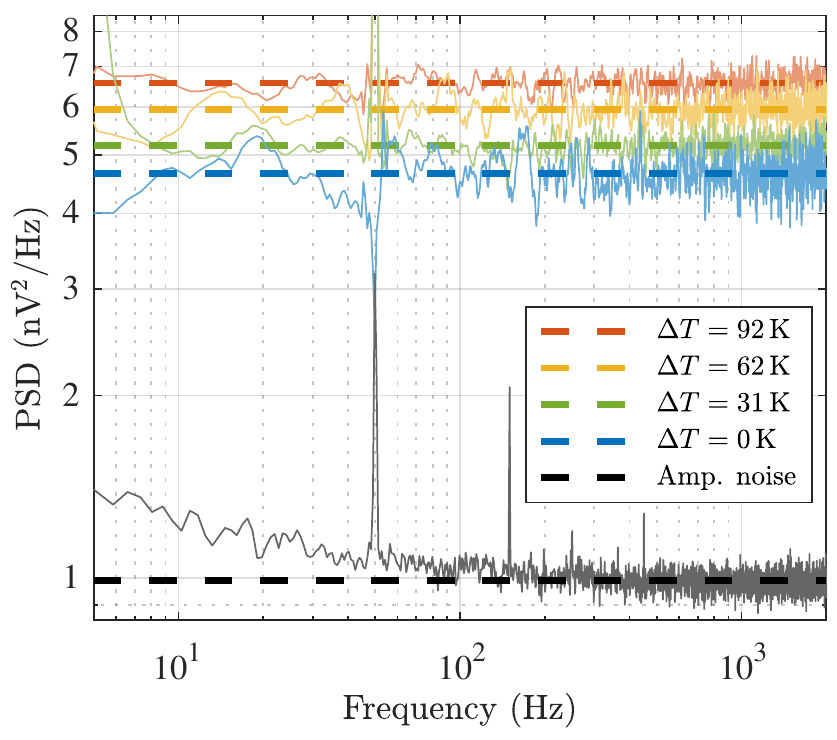}
\caption{Power Spectrum Density (PSD) of voltage fluctuations across the resistor for various $\Delta T$. The amplifier voltage noise (lower curve) has been subtracted from the displayed spectra. The noise is white from a few Hz to $\SI{2}{kHz}$, and increases with $\Delta T$. The thin lines correspond to the measured spectra, while the thick dashed lines are the mean value of the spectra on this frequency interval.} \label{fig:PSD}
\end{center}
\end{figure}

\begin{figure}[htbp] 
\begin{center}
\includegraphics{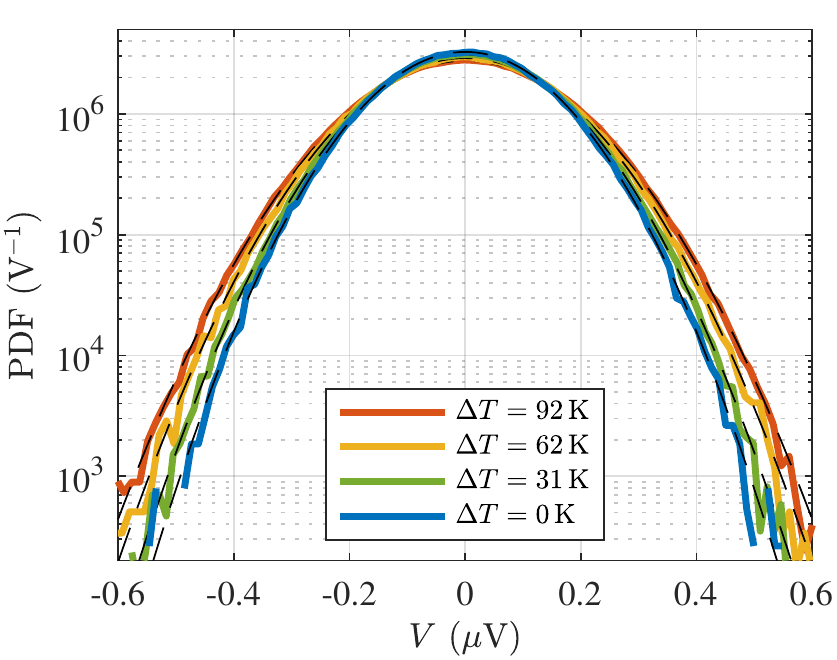}
\caption{Probability Distribution Function (PDF) of voltage fluctuations across the resistor for various $\Delta T$. Each PDF correspond to a minute long acquisition at $\SI{6}{kHz}$, filtered (AC + notch filter at $\SI{50}{Hz}$). The dashed lines are centered normal distribution fits of the experimental PDF: no deviation from gaussian noise can be detected. The variance of the signal increases with $\Delta T$.} \label{fig:pdfV}
\end{center}
\end{figure}

The statistical properties on the voltage fluctuations are illustrated in Figs.~\ref{fig:pdfV} and \ref{fig:PSD}, where we plot the Probability Distribution Function (PDF) of the signal, and its Power Spectrum Density (PSD).

In Fig. \ref{fig:PSD}, we plot the power spectrum density of the voltage signals, acquired during a minute long acquisition at $\SI{6}{kHz}$ for various $\Delta T$. The amplifier voltage noise has been subtracted from the displayed PSD. Except for a peak corresponding to line frequency, the spectra are flat in the $\SI{5}{Hz}$ to $\SI{2}{kHz}$ bandwidth displayed here. At higher frequencies, the amplifier input capacitance starts short-cutting the resistance and filtering the signal. At lower frequencies, the amplifier $1/f$ noise can be an issue. We therefore restrict ourselves to the displayed frequency range, and exclude the $\SI{50}{Hz}$ peak to compute the mean value of the plateau. The statistical uncertainty on its value is then around $\SI{0.3}{\%}$.

The statistical properties on the voltage fluctuations are also illustrated in Fig.~\ref{fig:pdfV}, where we plot the Probability Distribution Function (PDF) of the signal, for various $\Delta T$. To avoid artefacts from slow drifts in the temperature of the sample or to the amplifier noise (low frequency, line frequency), the PDF are computed from the signal digitally filtered with a high pass filter at $\SI{0.3}{Hz}$, and a notch filter with a $\SI{2}{Hz}$ stop band around $\SI{50}{Hz}$. For all observables and all probed $\Delta T$, the noise is perfectly white: its PDF is gaussian (skewness lower than $\num{6e-3}$, excess kurtosis lower than $\num{e-2}$), and its PSD is flat. Its variance, or its power spectrum density plateau, hence $\Tfluc$, is thus enough to describe the voltage fluctuations.

\begin{figure}[htbp] 
\begin{center}
\includegraphics{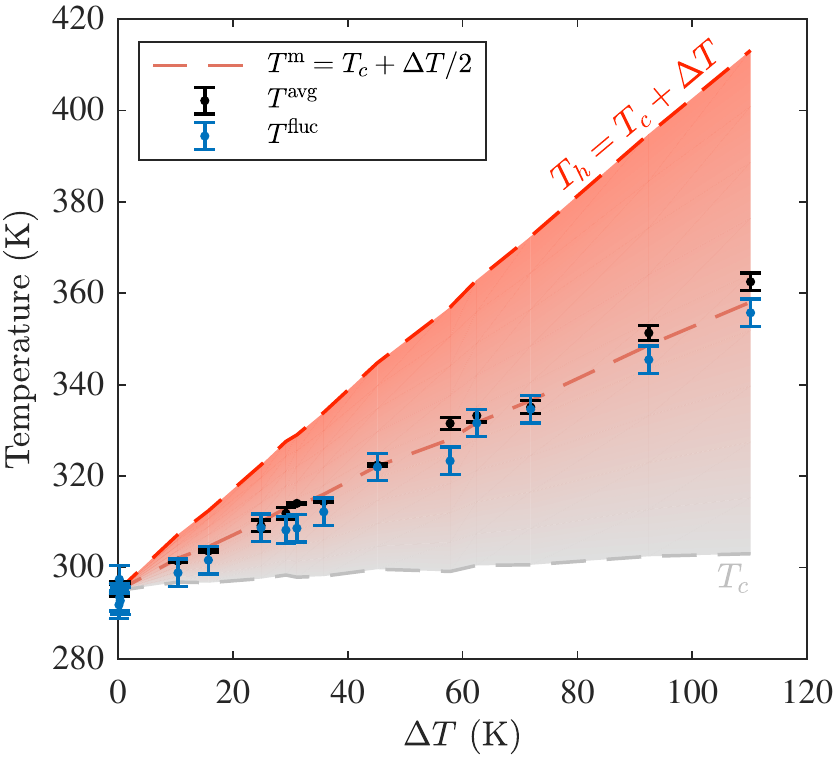}
\caption{Temperatures $\Tm$ (mean of the two thermostats), $\Tavg$ (spacial average) and $\Tfluc$ (from noise amplitude) as a function of the imposed temperature difference $\Delta T$. The proximity of the three measurements gives credit to the simple extension of the Nyquist formula in this NESS.} \label{fig:Tfluc}
\end{center}
\end{figure}

In figure \ref{fig:Tfluc}, we finally report the measured values for $\Tfluc$ as a function of $\Delta T$. We compare to the two estimation of the average temperature of the resistance $\Tm$ and $\Tavg$. Experimental uncertainties are around $\SI{1}{K}$: they correspond mainly to the statistical uncertainty for $\Tfluc$, to the uncertainty in $\alpha$ for $\Tavg$, and to temperature stability for $\Tm$. Within those error bars, all three temperatures stand very close, and give credit to the simple extension of the Nyquist Formula of eq.~\ref{eq:Tflucext}.

\section{Discussion and conclusions}

Taking into account the experimental results, to the first order in $\Delta T$, our approach seems very reasonable. Let us evaluate in the following lines if higher order effects could be present in our system.

The Wiedemann-Franz law for metals states that thermal and electrical conductivities are proportional~\cite{Franz-1853,Jones-1985}, which may be written in our case as:
\begin{equation} \label{eq:WiedemannFranz}
\rho(T) \lambda(T) =\rho_0 \lambda_0
\end{equation}
As illustrated by the measurements, the resistance has a noticeable variation on the temperature range we probe. Thus thermal conductivity present a similar variation, meaning that the temperature gradient is not uniform in space, even with negligible radiation losses. Evaluating $\Tfluc$ analytically from  eq.~\ref{eq:Tflucext} seems thus non trivial in the general case.

However, neglecting radiation, let us write that the heat flux is constant:
\begin{equation} \label{eq:jC}
J_\lambda = \frac{\pi}{4} D^2 \lambda(T) \partial_x T
\end{equation}
Combining eqs.~\ref{eq:WiedemannFranz} and \ref{eq:jC}, we have $\rho(T) = A \partial_x T$, where $A$ is a constant. Integrating this equation from $x=0$ to $L$, we immediately get $A = R/ \Delta T$, hence:
\begin{equation}
 \rho(T)  = \frac{R}{\Delta T} \partial_x T
\end{equation}
Let us now report this expression of the resistivity in the extended Nyquist prediction of eq.~\ref{eq:Tflucf}:
\begin{align}
\Tfluc_f	& = \frac{1}{R_f} \int_{x=0}^{l} T(x) \rho\big(T(x)\big) dx \\
				& = \frac{1}{\Delta T} \int_{x=0}^{l} T(x) \partial_x T(x) dx \\
				& = T_c + \frac{1}{2} \Delta T = \Tm
\end{align}
Under quite broad hypothesis (constant heat flux, Wiedemann–Franz law), the result is surprisingly simple: the amplitude of thermal noise corresponds to a temperature exactly at the mid-point between the two thermostats. Note that this result is valid even if the temperature field is non linear, and the resistivity non uniform.

In our experiment, this formula applies only to the free standings parts of the resistive wire, thus to $\Tfluc_f$. For the actual measurement, using eq.~\ref{eq:Tflucloop} and injecting the temperature dependence of $R_h=R_c(1+\alpha \Delta T)$, we get to the second order in $\Delta T$:
\begin{equation}
\Tfluc = T_c + \frac{1}{2} \Delta T + \frac{R_c}{4(R_c+R_f)} \alpha \Delta T^2
\end{equation}
In our configuration, the ratio $R_c/4(R_c+R_f)$ can be evaluated by $l_c/4(l_c+l_f)\approx 0.1$. The quadratic term is then only a $\SI{5}{K}$ correction at the end of the explored $\Delta T$ range. Other effects of similar amplitude but opposite direction (such as the contribution of the few centimeters of the resistive wire between the wrapped part and the connector to the amplifier, mainly at room temperature) make this contribution inobservable in our system.

To summarise our work, we have presented in this article an extension of the Nyquist formula to a resistance bridging two thermostats at different temperatures. On the basis of reasonable hypotheses we have analytically shown that the thermal noise in this NESS is equivalent to that of a resistance in equilibrium at the mean temperature between the two thermostats. Within an experimental accuracy of a few percent, we then demonstrated that this extended Nyquist Formula is valid in our setup. Beyond the results presented in this article, we probed a commercial resistor of $\SI{640}{k\Omega}$ (film resistor with ceramic substrate) and reached exactly the same conclusion. From a more general perspective, the thermal fluctuations are adequately described by the average of the temperature field weighted by the local dissipation. Beyond electrical observables, we expect this approach to hold in other expressions of the fluctuation-dissipation theorem.

\begin{acknowledgments}
We thank F. Vittoz and M. Tanaze for technical support, B. Huard and Q. Ficheux for stimulating discussions. This work has been performed using the low noise chamber of the Optolyse project, funded by the French r\'egion Auvergne-Rh\^one-Alpes (Optolyse, CPER2016).
\end{acknowledgments}

\bibliographystyle{unsrt}
\bibliography{ExtendedNyquist}

\end{document}